\documentclass[twocolumn,showpacs,preprintnumbers,amsmath,amssymb]{revtex4}
\usepackage[final]{graphicx}
\usepackage[english]{babel}
\begin{document}
\title{
The mesons and baryons production in nucleus-nucleus collisions at
SPS and RHIC energies and quark-gluon plasma }
\author{Yu.A. Tarasov}
\email{tarasov@dni.polyn.kiae.su} \affiliation{
 Russian Research
Center ''Kurchatov Institute'', 123182, Moscow,  Russia }
\date{\today}

\begin{abstract}
We use quasiparticle description of deconfined
matter in nuclear collisions at finite temperature and chemical
potential. We assume that evolution of expanding system is
isentropic. Using theoretical formulas and experimental meaning
for the average multiplicity of charged and neutral particles and
also of net nucleons ($N-\bar{N}$) per unit central rapidity, we
calculate the initial temperature $T_0$, and the volume $V_0$
 for the collisions of heavy nuclei (Pb+Pb and Au+Au)
at SPS and RHIC energies. At calculations we use the conservation
of entropy and of number net nucleons in initial plasma stage and
on the stage of phase transition at temperature $T_c$, where we
take into account the phase of constituent quarks $m_q$ and $m_s$.
 We use at calculation the effective quasiparticle model with
 decrease of thermal gluon mass $m_{g}(T)$ at $T\to T_c$ from
 above. This model agrees with lattice data.
 The particle ratios for SPS and RHIC are defined by early chemical
freeze-out at temperature close to $T_c$ (at values of chemical
potential $\mu_B=247$ MeV and 50 MeV for SPS and RHIC (at
$\sqrt{s}=130$ GeV) correspondingly). The spectra of baryons and
mesons are calculated at temperature of thermal freeze-out $T_f =
120$ MeV. The quantitative characteristic of these  spectra (for
example, the normalization for baryons and mesons) are determined
in the present model by initial state - quark-gluon plasma. At
calculation of spectra we assume that averaged transverse flow
velocity $\bar v_{\perp}$ have both direct particles and paternal
resonance. We show that values of $\bar v_{\perp}$ increase for
energy RHIC in comparison with SPS. We had investigated also the
analogous problem  of nuclear collisions in quasiparticle model
with phenomenological parametrization of running coupling $G(T)$,
which increase at $T\to T_c$ from above (in difference from
effective quasiparticle model). We have here agreement with new
lattice data. We do not found in this model quantitative
difference for spectra in comparison with effective quasiparticle
model. However it can be shown, that in this model we have
dramatic situation for energy loss of gluon jets in plasma.
\end{abstract}

 \maketitle

\section{Introduction    \label{sec1}  }

There is considerable interest to heavy ion reaction at
sufficiently high energies, because they provide the possibility
for conditions of the transitions to deconfined state of hadronic
matter-quark-gluon plasma (QGP). At Pb+Pb and Au+Au collisions
there is considerable number of net nucleons ($N-\bar{N}$) in
central interval of rapidity, which it is necessary take into
account on plasma and hadron stage. Therefore we have the baryon
chemical potential $\mu_B \ne 0$.

In the  works a number of authors (P.Braun-M\"{u}nzinger
et.al.~\cite{1}) was shown, that statistical thermal model gives
good description of particle ratios in central collisions of heavy
nuclei at high energies (for SPS and RHIC). The temperature $T$
for all this is close to critical temperature $T_c\approx 170$ MeV
obtained from lattice Monte-Carlo simulations of QCD, and $\mu_B$
is close to 50 MeV for RHIC(at $\sqrt{s}$=130 GeV)   This
correspond to chemical freeze-out, when inelastic collisions cease
to be important and the particle composition is fixed. The
trajectory of chemical freeze-out $T(\mu_B)$ was obtained also in
the work~\cite{2} on the basic of relation for hadronic gas
$<E_h>/<N_h>=const\approx 1$ GeV. For SPS was found the value
$\mu_B \approx 250-260$ MeV. One can to suppose that trajectory
$T(\mu_B)$ for various energies of nuclei is close to boundary of
deconfined quark-gluon plasma phase. For all this the potential
$\mu_s$ correspond to strangeness neutrality (i.e. overall
strangeness = 0). This correspond in fact to hadronic part of
mixed phase, where there are strange mesons and baryons (i.e.
$\mu_s\ne 0$).

In the present work we try to define the physical characteristic
of initial plasma phase $T_0$ and $V_0$. For this aim we use
hypothesis of conservation of entropy and of number of net
nucleons in initial stage (where $\mu_s = 0$) and on stage of
phase transition at $T =T_c$, where we take into account also of
constituent quarks phase.

The plasma at  finite temperature was considered perturbatively up
to order $O(g^5$)~\cite{3}. However in the experimentally
accessible region (close to temperature $T_c$ of phase transition)
the strong coupling constant is sufficiently large: $g\sim 2$. The
perturbative expansion in powers of g probably gives bad
convergence. Recently in series of papers~\cite{4,5,6} for
description of quark-gluon plasma have been used quasiparticle
model, which allow apparently to sum up partially the perturbative
expansion. At high temperature the plasma consist of quasifree
quarks and gluons and probably is described well by perturbative
theory.
The lattice model show rapid increase of entropy density
and other values near by $T_c$ in SU(3) gluon plasma. In
 quasiparticle model
 the interacting plasma of quarks and gluons is described as
system of massive quasiparticles.

For thermal momenta $k\sim T$ the quark-particle excitation and
transversal gluons will propagate with dispersion relations
\begin{eqnarray}
 \label{eq.1}
\omega^2_i(k)&\simeq& m^2_i(T) +k^2,
 \\
 m^2_i(T)&=&m^2_{0i} +\Pi_i, \nonumber
\end{eqnarray}
where $\Pi_i$ are given by the asymptotic values of the hard
thermal self energies~\cite{6,7}
\begin{eqnarray}
\label{eq.2}
 \Pi_q&=&2\omega_q(m_0+\omega_q),\; \omega_q^2=\frac{N_c^2-1}{16N_c}[T^2+\frac{\mu_q^2}{\pi^2}]G^2,    \\
 \Pi_g&=&
 \frac{1}{6}[(N_c+\frac{N_f}{2})T^2+\frac{6}{2\pi^2}\mu_q^2]G^2. \nonumber
 \end{eqnarray}

The contributions $\Pi_i$ are generated dynamically by the
interaction within the medium, and $m_i(T)$ acts as effective
mass. The quantity $G^2(T)$ is to be considered as effective
coupling in quasiparticles model. The entropy and energy density
for example for gluon plasma takes the form:
\begin{equation}\label{eq.3}
s_g(T)=\frac{g_g}{2\pi^2T}\int\limits_{0}^{\infty}
k^2f(E_k)\frac{4k^2+3m_g^2(T)}{3E_k}dk
\end{equation}
\begin{equation}\label{eq.4}
\epsilon_g(T)=\frac{g_g}{2\pi^2}\int\limits_{0}^{\infty}k^2f(E_k)(E_k)dk
+B(T),
\end{equation}
where $f(E_k)=[\exp(\frac{E_k}{T}) -1]^{-1}$,
$E_k=\sqrt{k^2+m_g^2(T)}$

The value $B(T)$, introduced in expression for energy (and also
for pressure) is necessary in order to maintain thermodynamic
consistency, that is must be fulfilled the relations $\epsilon +p
=sT$ and $s=\frac{\partial p}{\partial T}$. The function $B(T)$
compensates the additional $T$ derivatives  from  $m(T)$ in
pressure and it represent in integral form in the paper~\cite{5}.
In our paper we do not use the function $B(T)$.

However the phenomenological parameterization of coupling constant
$G^2$ in accordance with perturbative $QCD$ in quasiparticle
model~\cite{4,6} is very likely unjust close to temperature of
phase transition $T_c$, where it should be expected of
nonperturbative dynamics. In this model we have the increase of
thermal gluon mass in vicinity of phase transition.

In Sec.~\ref{sec2} we give the brief review of phenomenological
model of confinement~\cite{10}, which agrees with $SU(3)$ lattice
data. This model is applied also to system with dynamical quarks.

In Sec.~\ref{sec3} we calculate for isentropic evolution of
quasiparticles system with chemical potential $\mu_B \ne 0$ the
various physical characteristic: the initial temperature $T_0$,
the volume $V_0$,  the initial entropy $S_0$ etc. for nuclear
collisions at SPS and RHIC energies. We use the values $T_c$ and
$\mu_B$ close to above-mentioned: $T_c =170$ MeV and  \mbox{$\mu_B
= 50$ MeV and 247 MeV} for RHIC and SPS. At calculation we use
also additional information - experimental meaning of average
number of net nucleons in central region of rapidity for SPS and
RHIC. In result we have obtained the equations (\ref{eq.22}),
(\ref{eq.23}) for definition of $T_0$. Simultaneously must be
fulfilled the relations (\ref{eq.24}),which follow from
conservation of entropy and of number net nucleons.

Besides  of study of characteristic of initial phase we describe
in Sec.~\ref{sec3} another new aspect - we show, that from
conservation of entropy and of number of net nucleons follow, that
massive constituent quarks ($m_q$ and $m_s$) appears with decrease
of number of degrees of freedom (i.e. with effective number) in
the presence of octet of pseudogoldstone states. We show also,
that with the same effective number of degrees of freedom appears
hadrons and resonances in hadron part of mixed phase (for SPS and
RHIC).

Certainly it should be noted, that in nucleus-nucleus collisions
at high energies there is possibility of two stages of
equilibrium. The gluons reach of thermal equilibrium to a
considerable extent faster, than quarks. According to estimations,
equilibrium time of gluons is $\tau_g \sim 0.3$ fm/c and for
quarks one is $\tau_q \sim 2$ fm/c, i.e. it is possible the
production of hot glue at first stage ~\cite{11}. The
corresponding estimations of initial conditions for hot glue in
effective quasiparticle model there are in the work ~\cite{12}

In Sec.~\ref{sec3} we investigate the initial condition ($T_0$,
$V_0$) and evolution of second plasma stage, where is the
equilibrium also for quarks.
 In this paper we do not consider the
particle ratios. We have considered the ratios for various baryons
and mesons with accounting of resonance decays in the
work~\cite{33}. The particle ratios in the main agrees well with
experimental data for SPS and RHIC ($\sqrt{s}=130$ MeV.

In Sec.~\ref{sec.4} we calculate in effective quasiparticle model
the some baryons and mesons (pions and kaons) spectra with
accounting of resonance decays and transverse flow. We use for
heavy nuclei the thermal freeze-out temperature $T_f =120$ MeV.
The normalization of baryons and mesons spectra here is no free
parameter, but is defined by initial condition in plasma - by
initial entropy $S_0$ and by condition of entropy conservation
$S_f = S_0$. We show that values of averaged transverse flow
velocity $\bar v_{\perp}$ increase for RHIC in comparison with
SPS.

The main object of present work is no study the whole of spectra
for SPS and RHIC, but study of dependence spectra from initial
state and from stage of phase transition plasma into hadrons.

Therefore  in Sec.~\ref{sec5} we consider the analogous isentropic
problem of nuclear collisions by the use of ordinary perturbative
theory up to order $O(\alpha_{s}(T))$  with QCD parameter $\lambda
\sim 0.2$ GeV. We calculate here also the values $T_0$, $V_0$,
$S_0$. We do not find noticeable difference for spectra of
particles in comparison with effective quasiparticle model.
However this  model disagrees with SU(3) lattice data in region of
phase transition (i.e. near of temperature $T_c$).
 In Sec.~\ref{sec5} we consider also the quasiparticle model with
 phenomenological parametrization of running coupling $G(T)$,
 which gives good fit of new lattice data. We calculate here also
 the initial parameteres $T_0,V_0,S_0$. The coupling strength
 $G_T$ here increase at $T\to T_c$ from above (unlike effective
 quasiparticle model). However baryons and mesons spectra in this
 model practically do not differ from spectra in effective
 quasiparticle model. Thus the spectra weakly depend on region of
 phase transition.

However it can be shown~\cite{12}, that in ordinary peturbative
model (taking into account the hot glue) we have too great energy
loss $\Delta E_g$ of gluon jet at RHIC energy: $\sim 80-90 \%$
(unlike effective quasiparticle model). Thus the jet quenching
essentially depend on description of phase transition stage. We
show also, that in second above-mentioned model the situation with
energy loss $\Delta E_g$ is more dramatic.

In Sec.~\ref{sec7} - conclusion.

\section{Effective quasiparticle model    \label{sec2}}

In the papers~\cite{4,6} for gluon plasma was used the
phenomenological parameterization of coupling constant $G^2(T)$ in
accordance with perturbative QCD with the two fit parameters $T_s$
and $\lambda$. The resulting equation of state (EOS) is in good
agreement with lattice data over a wide temperature range between
$T_c$ and 5$T_c$. However in spite of this the thermal gluon mass
have strong increase in the vicinity of the phase transition at
decrease of $T$. Furthermore, close to a phase transition the
coupling constant $G_s(T_c)\sim 2$, that is correctness of
perturbative of calculations is questionable.

Further it is expected that correlation length $\xi(T)$, which is
proportional $m_D^{-1}$, grows when $T\to T_c$ from above ($m_D$
is gluonic Debye mass). The lattice calculations shows that $m_D$
drops by factor of 10 when $T$ decrease from 2 $T_c$ to
$T_c$~\cite{8} Also for three colors the value $\xi(T_c)$ remains
large, but finite. In hard thermal loop ($HTL$) perturbation
theory the thermal mass $m_g(T)$ and gluonic Debye mass $m_D$ are
connected by relation~\cite{9}

\begin{equation}\label{eq.5}
m_D =\sqrt{2}\;m_g.
\end{equation}

However in nonperturbative theory this relation can be broken
down~\cite{9}.

Recently in the paper~\cite{10} was considered interesting
phenomenological model of confinement. In region below $T_c$ in a
pure $SU(3)$ gauge theory are color singlet of heavy glueballs.
Approaching $T_c$ the gluons are liberated, followed by sudden
increase of energy and entropy density. When approaching the phase
transition from above, the decrease of thermodynamic values is not
caused by increase of masses(unlike~\cite{4,6}), but caused by
decrease of number active degrees of freedom. When $T$ comes
closer to $T_c$, more and more of gluons are absorbed by
glueballs. It is assumed that thermal gluons mass $m_g(T)$ follows
roughly the behavior of the Debye mass, i.e. it decreases. However
the entropy density $s(T)$ will overshoot the lattice entropy
because light masses near $T_c$. This difference may be accounted
in quasiparticle model by modifying of number effective degrees of
freedom in (\ref{eq.3}), (\ref{eq.4})
\begin{equation}\label{eq.6}
g_g \to C(T)g_g.
\end{equation}

The explicit value $C(T)$ may be estimated as the ratio of the
lattice entropy and entropy (\ref{eq.3}) with a dropping mass
$m_g(T)$.  At $T \gg T_c$ we expect $C(T) \simeq 1$ and near $T
\sim T_c$ we have $C(T)< 1$.  At $T < T_c$ it can estimate $C(T_c)
\sim 0.2$ from lattice data. The value $C(T)$ is a smooth,
monotonously increasing function with $T$.

It is assumed also that thermal mass (for example gluons mass)
have form $ m_g(T) =\tilde G(T) T$. The lattice results for Debye
mass $m_D(T)$ can be parameterized well by formula
\begin{equation}\label{eq.7}
 m_D(T)\simeq G_1 T (1+ \delta -\frac{T_c}{T})^\beta \equiv
 \left(\frac{N_{c}}{6}\right)^{1/2}\!\!\! G_{0}
T (1+ \delta - \frac{T_c}{T})^\beta
\end{equation}
where $G_1 \simeq 1.3, \beta \simeq 0.1$ and there is small gap at
$T=T_c$ with $\delta \simeq 10^{-6}$. The value $G_1$ is
determined here by the asymptotic value of thermal gluon mass,
which coincide with lattice mass for instance at $T = 3T_c$. Below
of this value the explicit $HTL$ resummation is expected to
fail~\cite{13}. One can to use the asymptotic value $G(T)$ with
 renormalization  group inspired parametrization ~\cite{6}:
 \begin{equation}\label{eq.8}
G^{2}(T,\mu=0)=\frac{24\pi^2}{(11N_c
-2N_f)\ln(\lambda(T+T_s)/T_c)}
\end{equation}
The parameteres $T_s$ and $\lambda$ are used for fitting new
lattice results with $N_f$=2 and $N_f$=4 at $T=3T_c$ (and also for
$N_f$=3 ~\cite{6}). The thermal gluon mass at chemical potential
$\mu_q$=0 is analogous with formula (\ref{eq.7}) where $G_1 \to
\sqrt{N_c/6 +N_f/12}G_0$. Effective coupling $G_0$ have the form
$G_0 = \frac{g_0}{\sqrt {11N_c - 2N_f}}$. The gluon and quark mass
coincide with lattice data at $g_0 \simeq$ 9.4 for $N_f$=2 and at
$g_0 \simeq$ 9.8 for $N_f$=3. That gives $G_0 \simeq$ 1.886 for
$N_f$=3. The like value $G_0$=1.9 we have in formula (\ref{eq.7}).
The such value $G_0\simeq 1.9$ we use in calculations. It should
be noted, that lattice results at finite chemical potential
$\mu_q$ point to greatly weak dependence of asymptotic thermal
masses from $\mu_q$~\cite{14}.

Since (in accord with assumption) the values $m_g(T)$ and $m_D(T)$
have similar trends near $T \sim T_c$, the mass $m_g(T)$ is
parameterized by analogy with formula (\ref{eq.7}). In order to
account for certainties, were investigated~\cite{10} some a range
of values $G_0$, $\beta$ and $\delta$.

A decreasing effective coupling strength $G(T)$ at \linebreak[4]
\mbox{$T \to T_c$} from above  can be understood, since at
decreasing of $T$ more and more gluons become confined and form
heavy glueballs. The effective glueball exchange interaction
between gluons are reduced. The total interaction can be
interpreted as a superposition nonperturbative multi\-gluon and
weak glueball exchange. The good fit~\cite{10} for $C(T)$
(\ref{eq.6}) have form similar with (\ref{eq.7})
\begin{equation}\label{eq.9}
C(T,T_c) = C_0 \left(1+\delta_c -\frac{T_c}{T}\right)^{\beta_c},
\end{equation}
where $C_0 \simeq$ 1.25, $\delta_c \simeq$ 0.0026, $\beta_c
\simeq$ 0.31.

The relations (\ref{eq.7}), (\ref{eq.9}) give good description of
$SU(3)$ lattice data for $\epsilon/T^{4}$, $3s/4T^{3}$, $3p/T^{4}$
for $T$ close to $T_c$.

Further it is possible to extrapolate  effective  quasiparticle
model to system with dynamical quarks. Unfortunately no lattice
data on thermal masses with dynamical quarks are available. We
assume that $N_c$ and $N_f$ dependence of $m_g$ and $m_q$ are
given by formulas (\ref{eq.2}), where effective coupling $G(T)$
have the form: \linebreak[4] \mbox{$G(T) = G_{0} (1+ \delta -
T_{c}/T)^{\beta}$}. For example, the thermal \linebreak[4] masses
of $u$ and $d$ quarks are:
\begin{eqnarray}\label{eq.10}
m_q^2&=& m_{q0}^2 + 2 m_{q0}G(T)\sqrt{\frac{N_c^2 -1}{16 N_c}(T^2
+\frac{\mu_q^2}{\pi^2})} +\\ \nonumber &+& 2\frac{N_ c^2 -1}{16
N_c}(T^2 +\frac{\mu_q^2}{\pi^2})G^2(T).
\end{eqnarray}

For $s$ quarks we replace $m_{q0}\to m_{s0}$, $\mu_q \to 0$. It is
assumed that parameterization of function $C(T,T_c)$ in the
presence of quarks and parameterization (\ref{eq.9}) for gluons
are similar, with some variation of parameters. For example, $C_0
= 1.25$, $\delta_{c} \simeq 0.02$, $\beta_{c}\simeq 0.28$ for two
light quarks~\cite{10}. The relations of type (\ref{eq.9}) for
$C(T,T_c)$ and relations
 of type (\ref{eq.10}) for $m_q$ and $m_g$ are inserted into formulas for entropy density of
quasiparticles.
 For $q(u,d)$ quarks we have the entropy density:
\begin{eqnarray}
\label{eq.11}
&&s_q(G_0,m_q,\mu_q,T,T_c) =  \\
&=& s_{q1}(G_0,m_q,\mu_q,T,T_c)+\nonumber \\
&+& s_{q1}(G_0,m_q,-\mu_q,T,T_c) -\nonumber \\
&-&\frac{\mu_q T^2 g_q}{2 \pi^2} (F_1(G_0,m_q,\mu_q,T,T_c)-\nonumber\\
&-& F_1(G_0,m_q,-\mu_q,T,T_c)) \nonumber
\end{eqnarray}
where
\begin{eqnarray}
&& s_{q1}(G_0,m_q,\mu_q,T,T_c) = \nonumber \\ &=& \!\!\!\frac{T^3
g_q}{2 \pi^2}\!\!\! \int\limits_0^\infty dx \frac{(4x^4 +3x^2
\frac{m_q^2(T)}{T^2})C(T,T_c)} {3\sqrt{x^2+\frac{m_q^2(T)}{T^2}}
[e^{\sqrt{x^2+\frac{m_q^2(T)}{T^2}}- \frac{\mu_q}{T}} +1]}
\nonumber
\end{eqnarray}
The formula (\ref{eq.11}) gives correct term already in order
$g^2$ at perturbative expansion.

For $s$ quarks $m_q\to m_s$, and $\mu_s=0$. For gluons the entropy
density is:
\begin{eqnarray}\label{eq.12}
&&s_g(G_0,m_g,T,T_c) =\\ \nonumber &=&\frac{T^3g_g}{2\pi^2}
\int\limits_0^\infty dx \frac{(4x^4 + 3x^2
\frac{m_g^2(T)}{T^2})C(T,T_c)}{3\sqrt{x^2+\frac{m_g^2(T)}{T^2}}
(e^{\sqrt{x^2+\frac{m_g^2(T)}{T^2}}} -1)}.
\end{eqnarray}
For the net nucleons density we have:
\begin{eqnarray}
\label{eq.13} &&n_1(G_0,m_q,\mu_q,T,T_c)= \\ \nonumber &=&
\frac{T^3}{6\pi^2} g_q [F_1(G_0,m_q,\mu_q,T,T_c)-
F_1(G_0,m_q,-\mu_q,T,T_c)], \nonumber
\end{eqnarray}
where
$$
 F_1(G_0,m_q,\mu_q,T,T_c) =
\int\limits_0^\infty dx \frac{x^2
C(T,T_c)}{e^{\sqrt{x^2+\frac{m_q^2(T)}{T^2}}-\frac{\mu_q}{T}} +1}.
$$
 We shall show below that in massive constituent quarks phase and hadron part of
mixed phase we have also decrease of the number of effective
degrees of freedom.

\section{ Investigation of physical characteristic of initial and
mixed phases   \label{sec3}}

In ''standard model'' of relativistic nuclear collisions after a
pre-equilibrium period the system becomes thermalised in form of
quark-gluon plasma with formation time $\tau_0$. The system then
expand, cools off and at the time $\tau_c$ it begins to hadronize.
At temperature $T_c$ appears the mixed phase --- at first plasma
part at $\tau_c$, then hadronic part of mixed phase at $\tau_H$.
At temperature near to $T_c \simeq 170$ MeV particles first
undergo a chemical freeze out, and then a thermal freeze out at
temperature $T_f \sim 100-140$ MeV. The particles ratio is defined
by temperature close to $T_c$. For $A-A$ collisions we have the
initial volume $V_0 =\pi R_A^2 \tau_0$, where $R_A \simeq 1.2
A^{1/3}$. We consider isentropic evolution. The entropy per unit
rapidity is conserved during the expansion~\cite{16,17}. The value
$\frac{dS}{dy}=\pi R_A^2 s(\tau) \tau = const
(\frac{dN}{dy})_{y=0}$ determine initial entropy density, where
$\frac{dN}{dy}$ is the average multiplicity charged plus neutral
particles per unit central rapidity. In this case part of initial
energy goes into collective motion of the expanding system, that
is the initial energy density must have been higher than  in the
free flow scenario. In the work~\cite{18} was investigated the
isentropic longitudinal expansion of ideal gas of gluons and three
flavors of massless quarks. In this work was found the dependence
$\epsilon_0 \simeq (\frac{dN}{dy})^{4/3}$ for initial energy
density in plasma.
   In this paper we have investigated more common case for
quasiparticles with $\mu_B \ne 0$. The quasiparticles density in
plasma phase is:
\begin{equation}\label{eq.14}
n_0 = n_q +n_s +n_g,
\end{equation}
where $n_q,n_s,n_g$ is density of $u,d,s$ quarks and gluons. We
have here:
\begin{eqnarray}\label{eq.15}
&&n_q(G_0,m_q,\mu_q,T_0T_c) \\\nonumber&=&\frac{g_q T_0^3}{2\pi^2}
[F_1(G_0,m_q,\mu_q,T_0,T_c) + F_1(G_0,m_q,-\mu_q,T_0,T_c)],
\nonumber
\end{eqnarray}
\begin{equation}\label{eq.16}
n_s(G_0,m_s,T_0,T_c) = \frac{g_s T_0^3}{2\pi^2}
\int\limits_0^\infty dx \frac{x^2 C(T_0,T_c)}{e^{\sqrt{x^2 +
\frac{m_s^2(T_0)}{T_0^2}}} +1},
\end{equation}

\begin{equation}\label{eq.17}
n_g(G_0,m_g,T_0,T_c) =\frac{g_g T_0^3}{2\pi^2}
\int\limits_0^\infty dx \frac{x^2 C(T_0,T_c)}{e^{\sqrt{x^2
+\frac{m_g^2(T_0)}{T_0^2}}} -1}.
\end{equation}

By analogy it can be written for summary entropy density:
\begin{equation}\label{eq.18}
s_0 = s_q + s_s +s_g,
\end{equation}
where $s_q,s_s,s_g$ given by (\ref{eq.11}), (\ref{eq.12}). We must
express the initial temperature $T_0$ across the number of
secondary particles per unit central rapidity. For convenience of
calculations we introduce the designation: $n_0\equiv T_0^3
n_{01}$, $s_0\equiv T_0^3 s_{01}$, $\epsilon_0\equiv
T_0^{4}\epsilon_{01}$, that is we single out the factors $T_0^3$,
$T_0^4$. We single out also the constant factors $n_{01}\equiv 12
n_{02}/(2\pi^2)$ , $\epsilon_{01}\equiv 12\epsilon)_{02}/(2\pi^2)$
Thus we have $T_0 = (n_{0}/n_{01})^{1/3}$, $\epsilon_{0} =
(n_{0}/n_{01})^{4/3} 12\epsilon_{02}/(2\pi^2)$. Hence we have here
$\epsilon_{0}\equiv\left(\frac{N}{V_0}\right)^{4/3} f$, where $f =
\frac{\epsilon_{02}(2\pi^{2}/12)^{1/3}}{(n_{02})^4/3}$ .

Thus the initial temperature $T_0$ have been expressed across the
number of secondary per unit central rapidity:
\begin{equation}\label{eq.19}
T_0 = \left(\frac{dN}{dy
V_0}\right)^{\frac{1}{3}}\left(\frac{2\pi^2}{12n_{02}}\right)^{\frac{1}{3}}.
\end{equation}
For $A-A$ collisions the value $\frac{dN}{dy}$ can be extrapolated
by the form~\cite{19}:
\begin{equation}\label{eq.20}
\left(\frac{dN}{dy}\right)_{AA} = \left(\frac{dN}{dy}\right)_{pp}
A^{\alpha}.
\end{equation}
For $pp$ collisions we have good approximation per unit central
rapidity~\cite{20}:
\begin{equation}\label{eq.21}
\left(\frac{dN}{dy}\right)_{y=0}\simeq 0.8 \ln\sqrt{s}.
\end{equation}

The value $\alpha$ describes the amount of rescattering in the
interaction of the nuclei. From $p-A$ data it can expect
$\alpha\simeq 1.1$. The value $\alpha$ = 1.1 describes the result
for various light ion quite well~\cite{19}. The data for SPS
($\sqrt{s}=17.2$ GeV) is $\alpha = 1.08\pm 0.06$ (i.e. close to
($p-A$) data).  We have \mbox{$\frac{dN}{dy}\simeq 803$} (for
$\alpha$ =1.1) The data of Phenix collaboration show increase of
$\frac{dN_{ch}}{dy}$ from SPS (Na49) to RHIC ($\sqrt{s}=130$ GeV)
on value $\simeq$ 1.7, i.e. on value
 $\frac{\ln(\sqrt{s}=130)}{\ln(\sqrt{s} =17.2)}$. For SPS we assume $\alpha = 1.1$.
 From formula (\ref{eq.20}) we have for RHIC ($A=197$): $\alpha\simeq 1.11$

The initial volume $V_0$ in formula (\ref{eq.19}) can be expressed
across experimental meaning of average number of net nucleons
$N-\bar N$ in central region of rapidity. For SPS the number of
net protons in central region of rapidity is estimated as
28-29~\cite{21}. We use the value $N-\bar N = 57$.  The number of
net protons for RHIC in central region of rapidity (at
$\sqrt{s}=130$ GeV) is estimated as 8-10~\cite{22}, i.e. the
number of net nucleons is $\sim$ 16-20. We use the value $N-\bar N
\simeq$ 16.3 (for some greater agreement with spectra).

We have: $N-\bar N =V_0 \; n_1(G_0,m_q,\mu_q,T,T_c)$ (from
(\ref{eq.13}). Hence the value $T_0$ is:
\begin{equation}\label{eq.22}
T_0 =\left(\frac{dN}{dy} \frac{n_1}{N-\bar N}\right)^{1/3}
\left(\frac{2\pi^2}{12 n_{02}}\right)^{1/3}.
\end{equation}
We single out here the factor $T_0^3$ in formula (\ref{eq.13}) for
$n_1$ and take into account the values $\frac{dN}{dy} \simeq$ 803
and $N-\bar N \simeq$ 57 for SPS and $\frac{dN}{dy} \simeq$ 1374
and$N-\bar N \simeq$ 16.3 for RHIC.

 In result we have the equation for definition of initial
temperature $T_0$:
\begin{equation}\label{eq.23}
D_0 \equiv a_0
\left(\frac{d_{2}(G_0,m_q,\mu_q,T_0,T_c)}{n_{02}(G_0,m_q,\mu_q,T_0,T_c,m_s)}\right)^{1/3}
=1,
\end{equation}
where
\begin{eqnarray}
&& d_2(G_0,m_q,\mu_q,T_0,T_c) \equiv \nonumber \\ &\equiv&
(F_1(G_0,m_q,\mu_q,T_0,T_c) - F_1(G_0,m_q,-\mu_q,T_0,T_c)).
\nonumber
\end{eqnarray}
For calculations of values $n_{02}$ we use formulas
(\ref{eq.14}-\ref{eq.17})

For SPS we have with accounting of constant factors: $a_0$ = 1.676
at $\alpha =1.1$ and for RHIC $a_0$ = 3.045
 at $\alpha$ = 1.11.
We have from here the estimations of initial temperature $T_0
\simeq 175$ MeV for SPS and $T_0 \simeq$ 219.6 MeV for RHIC.

We use here the meaning for temperature of phase transition $T_c =
170$ MeV, as indicated by result from lattice gauge
theory~\cite{23}, and close to $T_c$ the temperature of chemical
freeze-out. At SPS energies we have for value $\mu_{B} = 247$ MeV
the good approximation for different antiparticles and particles
ratios. For RHIC ($\sqrt{s}=130$ GeV) we use also $T_c = 170$ MeV
and $\mu_{B} = 50$ MeV. That give the ratio $\bar p/p \approx
0.56$ close to found before experimental value. Later was
given~\cite{24} the ratio $\bar p/p =0.6 \pm 0.04 \pm 0.06$. In
the work~\cite{1} were given more precise values $\mu_{B} = 46$
MeV, $T_c =174$ MeV and particle ratios(at $\sqrt{s}=130$ GeV).
However there are for the present noticeable statistic and
systematic mistake in data. One should that not great distinction
have weak influence for example on spectra baryons and mesons.

 We use  also  the conservation of
entropy and of number net nucleons in initial phase at $T = T_0, V
= V_0$ and in mixed phase at $T = T_c, V = V_c$ (the entropy
density is $s_c$): $s_0(T_0) V_0 = s_c(T_c) V_c$, $n_1(T_0) V_0 =
n_1(T_c) V_c = N_{net} = N -\bar N$, i.e. $V_0 = \frac{N -\bar
N}{n_1(T_0)}$, $V_c = \frac{N -\bar N}{n_{1}(T_c)}$.

The entropy density $s_c(T_c)$ correspond to sum (\ref{eq.18}),
where $T_0 \to T_c$. Hence we have relations:
\begin{equation}\label{eq.24}
\frac{s_0(T_0)}{n_1(T_0)} = \frac{s_c(T_c)}{n_1(T_c)}.
\end{equation}

These relations depend also from constant $G_0$. The relations
 (\ref{eq.23}) and (\ref{eq.24}) must be fulfilled simultaneously.
   For SPS ($\mu_B$ =247 MeV,
 $T_c =170$ MeV) these relations indeed are fulfilled. It can be seen from
 mentioned below values. For
example for $\alpha = 1.1$ at $G_0 \simeq 1.9$, $T_0 \simeq 175$
MeV we have : $\frac{s_0(T_0)}{n_1(T_0)}=\frac{s_c(T_c)}{n_1(T_c)}
= 61.4$. It should be noted, that relations (\ref{eq.24}) are
fulfilled only for coincides values $\mu_B$ at initial and mixed
phases.   Thus, in this model the initial temperature $T_0$ at SPS
energy exceed weakly the temperature $T_c$ of phase transition. We
will express all physical values in units \linebreak[4]
\mbox{$m_{\pi} = 139$ MeV} $\simeq \frac{1}{1.42}$ fm$^{-1}$.

From  formulas (\ref{eq.11})-(\ref{eq.13}) - we find the values of
entropy density for SPS at $T_0$ and $T_c$: $s_0(T_0)\simeq 19.43
m_{\pi}^3$, $s_c(T_c)\simeq 15.31 m_{\pi}^3$  and also net
nucleons density $n_1(T_0)=0.314 m_{\pi}^3$, $n_1(T_c)=0.25
m_{\pi}^3$.

 For the number of net nucleons \linebreak[4]\mbox{$(N - \bar N)
= n_1(T_0) V_0\simeq 57$} we find the volume of initial plasma
phase: $V_0\simeq 181.3 m_{\pi}^{-3}\simeq 518$ fm$^3$ (i.e.
$\tau_0 \simeq 3.28$ fm). The initial entropy is: $S_0 = s_0 V_0
\simeq 3508$. We have also the volume $V_c$ of plasma part of
mixed phase, taking into account the conservation of entropy and
of net nucleons: $V_c = \frac{s_0 V_0}{s_c(T_c)} = \frac{N -\bar
N}{n_1(T_c)}\simeq 228 m_{\pi}^{-3} \simeq 650$ fm$^3$ (i.e.
$\tau_c \simeq 4.1$ fm).

Therefore with accounting of valent $u,d$ quarks we have the large
initial $\tau_0 > 1$ fm. The suitable calculations by formula
(\ref{eq.19}) for $\tau_0 = 1$ fm (for SPS) shows that for various
values of $G_0$ the number 57 of net nucleons correspond
$\mu_B\simeq 330$ MeV, that contradict to experimental data for
particles ratios. For correct values of $\mu_B \sim 200-250$ MeV
the number $N -\bar N$ is too little($\sim (35 - 45)$). The
analogous situation there is and for RHIC.

At decrease of interaction the thermal masses of quasiparticles
$m_{q}(T)$ and
 $m_{s}(T)$ (\ref{eq.10}) are diminished at $T\to T_c$ from above, but they do
not correspond to constituent masses of quarks in mixed phase. For
example, in the papers~\cite{25,26} was shown, that by means of
help of masses constituent quarks $m_u =m_d = 363$ MeV, $m_s =
538$ MeV can be found the masses of baryons in octet. For $SU(3)$
group the spontaneous breaking of chiral symmetry and appearance
of masses leads to appearance of octet pseudoscalar light
Goldstone (or rather pseudogoldstone) states $\pi,k$,
$\eta$~\cite{27}. However if in place of relation
$\frac{s(T_c)}{n_1(T_c)}$ in ~(\ref{eq.24})
 to use relations for
massive quarks with accounting of pseudogoldstone states, then
these relations  do not fulfilled. But these relations will be
satisfied, if to use decrease of the number of effective degrees
of freedom of massive quarks. We calculate the entropy  and
 net nucleon density $s_c$ and $n_c$ of massive quarks by formulas for ideal gas. For
example for $m_u =363$, $m_s =538$ MeV at $\mu_B =247$ MeV, $T_c =
170$ MeV we have $s_{c}(T_c) = 14.77 m_{\pi}^3$, and $n_{c}(T_c) =
0.33 m_{\pi}^3$.

For entropy density $s_{ps}$ of pseudogoldstone states $s_{ps} =
s_{\pi} + s_{k} + s_{\eta}$ we find by same way: $s_{ps}\simeq
3.76 m_{\pi}^3$ (at $T_c=170$ MeV, $m_{\pi}\simeq 135$ MeV). The
factor $\beta_{1}$ of decrease of number degrees of freedom can be
defined from relation \linebreak[4] \mbox{$\frac{s_{c}\beta_1 +
s_{ps}}{n_{c}\beta_1} \simeq 61.4$}. That gives $\beta_1 \simeq
0.685$.

Therefore, it is possible to interpret, that there is appearance
of massive constituent quarks with effective number of degrees of
freedom in the presence of octet of pseudogoldstone states, while
the multigluon state is absorbed into constituent quarks masses.

Let us list appropriate calculations for RHIC \linebreak[4]
(\mbox{$\sqrt{s} =130$} GeV). For example, at $\alpha = 1.11$
(\ref{eq.20}), $G_0 =1.9$, \mbox{$T_c =170$ MeV} we find from
(\ref{eq.23}): \linebreak[4] \mbox{$T_0 \simeq 219.6$ MeV}, that
is from here: $s_0(T_0) = 56.44$, $s_c(T_c)= 14.8$, \linebreak[4]
\mbox{$n_1(T_0) = 0.145$}, $n_1(T_c) =0.048$ in units of
$m_{\pi}^3$.
 The initial volume $V_0 \simeq \frac{16.3}{n_1(T_0)} \simeq 112.5 m_{\pi}^{-3} \simeq 322.5$ fm$^3$,
that is $\tau_0\simeq 2.2$ fm (for $\pi (R_{Au})^2 \simeq 148$
fm$^2$). The initial entropy
 $S_0 = s_0(T_0) V_0 \simeq 6350$.

 Hence we have for RHIC
$\frac{s_0(T_0)}{n_1(T_0)}\simeq 389$, $\frac{s_c(T_c)}{n_1(T_c)}
\simeq 308$, that is the relations (\ref{eq.24}) do not fulfilled.
However one can assume that at sufficiently high initial
temperature \mbox{$T_0 > T_c$} it is possible already the
appearance (in nonperturbative state at $T = T_c$) of
pseudogoldstone  with entropy density $s_{ps}$. It gives in fact
$\frac{s(T_c) + s_{ps}}{n_1(T_c)} \simeq 387$. That is we have on
this stage of mixed phase for RHIC $s_0 V_0 \simeq
(s(T_c)+s_{ps})V_c$,
 where $V_{c} = \frac{N - \bar N}{n_1(T_c)} \simeq 339
m_{\pi}^{-3}$, i.e. $\tau_c \simeq$ 6.5 fm.

At high energy (RHIC) the multigluon state is also absorbed into
massive constituent quarks with lesser number of degrees of
freedom. Can be found the factor $\beta_{3}$ of decrease of
effective number of degrees of freedom. We calculate here by
analogy with SPS the entropy density $s_{c1}$ and net nucleon
density $n_{c1}$ for massive quarks \mbox{$m_{u,d} = 363$ MeV},
\linebreak[4] \mbox{$m_s =538$ MeV} at $\mu_{B} = 50$ MeV, $T_c =
170$ MeV. We find in result: $s_{c1}=14.09 m_{\pi}^3$ and
$n_{c1}=0.0643 m_{\pi}^3$   We must have ratio $\frac{s_{c1}
\beta_3 +s_{ps}}{n_{c1} \beta_3} \simeq$ 390. We find from here
$\beta_3 \simeq$ 0.343. We have effective volume $V_{c1}$ on the
stage of constituent quarks \mbox{$V_c = \frac{S_{0} V_{0}}{s_{c1}
\beta_{3} +s_{ps}} \simeq 739 m_{\pi}^{-3}$}, or
\linebreak[4]\mbox{$V_c = \frac{N - \bar N}{n_{c1}(T_c)\beta_3}
\simeq 739 m_{\pi}^{-3}$} - is the same.

Let us consider now the hadronic part of mixed phase. We have
seen, that entropy density of massive constituent quarks is
defined by effective number of degrees of freedom and by
contribution of octet pseudogoldstone . It can be shown that with
such effective number of degrees of freedom ($\beta_1$ and
$\beta_3$) in hadron part of mixed phase appears nucleons and
hadrons. In order to find strange chemical potential $\mu_s$ for
hadron part of mixed phase, we use condition of disappearance of
strangeness:
\begin{equation}\label{eq.25}
\sum_{i}n_s^i - \sum_{i}n_{\bar s}^i = 0
\end{equation}
that is: $\sum_{i}g_{i}(F_{1i} - \bar F_{1i}) -
\sum_{j}g_{j}(F_{2j} -\bar F_{2j})s = 0$. Here
$$
F_{1i} = \int\limits_0^{\infty}dx \frac{x^2}{e^{\sqrt{x^2 +
\frac{m_i^2}{T^2}} - \frac{\mu_s}{T}} - 1},
$$
$$
\bar F_{1i} = F_{1i}(\mu_s \to - \mu_s),
$$
$$
F_{2j} = \int\limits_0^{\infty}dx \frac{x^2}{e^{\sqrt{x^2 +
\frac{m_j^2}{T^2}} - \frac{\mu_B - s \mu_s}{T}} + 1},
$$
$$
\bar F_{2j} = F_{2j}({\mu_B - s\mu_s} \to {-\mu_B + s\mu_s}).
$$

For strange baryons we have $s = 1$ except $\Xi (s = 2)$ and
$\Omega (s = 3)$. At temperature $T_c$ one should take into
account the considerable number of resonances. We took into
account the strange mesons up to $m_k^* = 1820$ MeV and the
strange baryons up to $m_B = 1940$ MeV. In result of calculatios
we have from (\ref{eq.25}): $\mu_s\simeq 61.44$ MeV at $\mu_B =
247$ MeV and $\mu_s\simeq 11.23$ MeV at $\mu_B = 50$ MeV (for $T_c
= 170$ MeV).

At SPS we find the entropy density according to calculation by
formulas of ideal gas: $\simeq 5.44 m_{\pi}^3$ for nonstrange
mesons and $\simeq 3.26 m_{\pi}^3$ for strange ones (at $T_c$).
For summary entropy density of strange baryons and the whole of
nonstrange ones \linebreak[4]\mbox{$n + p + \Delta + N_{440}^{*}
+\cdots$ we find: $\simeq 5.82 m_{\pi}^3$}, i.e. the whole entropy
is: $s_H \simeq 14.52 m_{\pi}^3$. For net nucleons density with
accounting of nucleonic resonances $N^{*}$ the calculation gives:
$n_H \simeq 0.402 m_{\pi}^3$. Hence the ratio is: $\frac{s_H}{n_H}
\simeq 36.2$, i.e. $\frac{s_H}{n_H} < \frac{s_c}{n_c}$. Taking
into account only nucleons(without of resonances), we have
$\frac{s_H}{n_H} = \frac{11.5}{0.1316}\simeq 87.3 >
\frac{s_c}{n_c}$. For $N + \Delta$ we have $\frac{s_H}{n_H}\simeq
46.5$, for $N + \Delta + N_{1440}^{*} + N_{1520}^{*} +
N_{1535}^{*}$ the calculation gives $s_H = 13.19 m_{\pi}^3$, $n_H
=0.303 m_{\pi}^3$, i.e.
 $\frac{s_H}{n_H}\simeq 44$ and so on. Thus  always
$\frac{s_H}{n_H} \ne \frac{s_c}{n_c}$, i.e. the relations
(\ref{eq.24}) do not fulfilled.

However for effective number of nucleon resonances $N + \Delta +
N_{1440,1520,1535}^{*}$ with lesser number of degrees of freedom
by factor $\beta_{1} = 0.685$ in the presence of octet
pseudogoldstones we have the ratio: $\frac{13.19 \beta_1 +
3.76}{0.303 \beta_1} = \frac{s_H^{eff}}{n_H^{eff}} \simeq 61.4$,
i.e. the relations (\ref{eq.24}) are fulfilled.
  We have also the volume $V_H^{eff} =
\frac{s_0 V_0}{s_H^{eff}}\simeq 274 m_{\pi}^{-3}$, or the same
$V_H^{eff} = \frac{N - \bar N}{n_H^{eff}} = \simeq 274
m_{\pi}^{-3}$, where $V_H^{eff}$ is effective volume of hadron
part of mixed phase. We give now the result of calculations at
RHIC energy.  The calculation gives for nucleons only (without
accounting of resonances):\linebreak[4]\mbox{$\frac{s_H}{n_H} =
\frac{10.335}{0.0195}\simeq 530$}, for $n + \Delta$ this ratio is:
$\simeq 282$, for the whole of nucleon resonance we have for this
ratio: $\simeq 205$ and so on, i.e. $\frac{s_H}{n_H} \ne
\frac{s_c}{n_c}$. But for the whole resonances the calculation
gives: $s_H =11.85$, $n_H = 0.0585$ in units $m_{\pi}^3$. With
lesser number of degrees of freedom by value $\beta_3$ =0.343(like
for massive quarks) we have $\frac{s_H^{eff}}{n_H^{eff}} =
\frac{11.85 \beta_3 +3.76}{0.0585 \beta_3} \simeq 390$. Now the
effective volume is: $V_H^{eff} \simeq \frac{16.3}{n_H^{eff}}
\simeq 812 m_{\pi}^{-3}$, or the same $\frac{s_0 V_0}{s_H^{eff}}
\simeq 811 m_{\pi}^{-3}$.

Hence the relations of type (\ref{eq.24}) are fulfilled also for
ratio $\frac{s_H}{n_H}$ in hadron part of mixed but with effective
number of degrees of freedom the same as for constituent quarks.

\section{The baryons and mesons spectra \label{sec.4}}

We  investigate  the spectra for SPS and RHIC at thermal
freeze-out. Apparently the favor results for heavy nuclei (for
example Pb + Pb) correspond to thermal freeze-out temperature $T_f
\approx$ 120 MeV~\cite{31,32}
 We consider at first SPS energy.
 We find the volume $V_f$ at \linebreak[4] \mbox{$T_f = 120$ MeV} from
conservation of number net nucleons. The value $\mu_B^f$ at
thermal freezing we find from condition: $\frac{\bar
p}{p}(T_c)=\frac{\bar p}{p}(T_f)$, where $ \frac{\bar
p}{p}(T_c)=e^{-2\mu_B/T_c}$. From here we have $\mu_B^f \simeq
174.35$ MeV. From formula (\ref{eq.13}) we have net nucleons
density at $T_f =120$ MeV, $\mu_B^f =174.35$ MeV: $n_f^{N-\bar N}
= 7.186 \times 10^{-3} m_{\pi}^3$. For $N_{net}\simeq 57$ we have
$V_f = N_{net}/n_f \simeq 7950 m_{\pi}^{-3}$. The value $\mu_s^f$
at 120 MeV may be found by no condition (\ref{eq.25}) (as the
chemical equilibrium already is absent), but from ratio:
$\frac{\bar \Lambda}{\Lambda}(T_c)= \frac{\bar
 \Lambda}{\Lambda}(T_f)$  From here we have found with accounting of
various weak decays: $\mu_s^f \simeq 40-42$ MeV. From relations
$\frac{n_{k+}}{n_{k-}}(T_c)=\frac{n_{k_+}}{n_{k-}}(T_f)$ (with
accounting of decays) we find for SPS the near value: $\mu_s^f$=41
MeV~\cite{33}. One can now find by formulas of ideal gas the
entropy density (at $T_f$) of nucleons and strange baryons: $s_B^f
\simeq 0.125 m_{\pi}^3$ and for nonstrange and strange mesons up
to $m_i$=1820 MeV: $s_{h+k}^f\simeq 1.3 m_{\pi}^3$. Thus the
entropy of baryons is \mbox{$S_B^f = V_f
 s_B^f \simeq 992$}, and the entropy of mesons \linebreak[4] \mbox{$S_{h+k}^f = S_0
- S_B^f \simeq 2516$}. The volume of mesons at thermal freeze-out
is $V_{h+k}^f = \frac{S_{h+k}^f}{s_{h+k}^f} \simeq 1940
m_{\pi}^{-3}$. We can estimate the number of $\pi_0$:
 \mbox{$N_{\pi_0} = V_{h+k}^f  n_{\pi_0}$}.For of the value
\linebreak[4] \mbox{$n_{\pi_0^f} \simeq 0.0807$}
 (calculated without accounting of weak decays) we have  \linebreak[4] \mbox{$N_{\pi_0}^f \simeq 157$}. The
experiment per unit of rapidity gives \linebreak[4]
\mbox{$N_{\pi_0} = 165 \pm 20$~\cite{34}}. By analogy we find also
the number of $k_{+}$: \linebreak[4] \mbox{$N_{k_{+}} = n_{k_{+}}
V_{h+k}^f \simeq 0.0149 V_{h+k}^f \simeq 29$}. Thus in this model
we have different volume of freeze-out for baryons and mesons. The
spectra of protons and strange baryons one should to calculate
with $V_f$, and of mesons
--- with volume $V_{h+k}^f$.

The relations of type (\ref{eq.24}) must be fulfilled also for
thermal freeze-out. Here we have $S_0 = S_f = s_{h+k}^{f}
V_{h+k}^f +s_{B}^{f} V_f$, and for net nucleons: $n_{1} V_0 =
n_{f} V_f \simeq 57$, i.e. we have instead of (\ref{eq.24}) the
relation: \linebreak[4] \mbox{$\frac{s_{0}(T_0)}{n_{1}(T_0)} =
\frac{s_{h+k}^{f} V_{h+k}^{f}}{n_{f} V_f} + \frac{s_B^f}{n_f}
\simeq 44 + 17.4 =61.4$}. Thus the relations of type (\ref{eq.24})
are fulfilled (but now without of decrease of number of degrees of
freedom) And backwards from  conservation of entropy $S_0 = S_f$
one can to find the final volume $V_f$ for baryons.

We calculate the spectra $p - \bar p$, $\Lambda$, $\Xi$, $\Omega$
direct and with resonance decays. We take into account the
transverse flow. We use the hydrodynamic model with linear
transverse velocity profile $v_{\perp}(r/R)\equiv v_{\perp}(x) =
\frac{3}{2} \bar v_{\perp}  x$, where $\bar v_{\perp}$ is the
averaged transverse flow velocity. The spectra of net protons
expressed by formula:
\begin{eqnarray}
\label{eq.26} \frac{dN^{p-\bar p}}{m_{\perp}dm_{\perp}} &=& V_f
\frac{(e^{\mu_f/T_f} - e^{-\mu_f/T_f})}{\pi^2} m_{\perp} g_p
\times \\ &\times& \int\limits_0^1 dx x I_0(
p_{\perp}\frac{\sinh(\rho)}{T_f}) K_1(
m_{\perp}\frac{\cosh(\rho)}{T_f}).\nonumber
\end{eqnarray}
 Here $\cosh(\rho) = \frac{1}{\sqrt{1 - v_{\perp}^2(x)}}$,
$\sinh(\rho) = \frac{v_{\perp}(x)}{\sqrt{1 - v_{\perp}^2(x)}}$.

For $\Lambda, \Xi, \Omega$ we have factors $\exp(\frac{\mu_B -n
\mu_s}{T_f})$, where \linebreak[4] \mbox{$n = 1,2,3$}
correspondingly, and for $k_{+}$ - factor \linebreak[4]
\mbox{$V_{h+k}^f \exp(\frac{\mu_s}{T_f})$}. We take into account
also spectra at resonance decays, for example: \mbox{$\Xi^0 \to
\Lambda\pi^0(99.54 \%)$}, \linebreak[4] \mbox{$\Xi_{-} \to
\Lambda\pi^{-}(99.88\%)$}, \mbox{$\Xi_{1530} \to \Xi\pi$},
\mbox{$\Sigma_{1190} \to \Lambda\gamma$}, \linebreak[4]
\mbox{$\Sigma_{1385} \to \Lambda\pi$}, $\Omega^{-} \to
\Xi^0\pi{-}$.

At resonance decays we must take into account the resonance with
transverse flow, and cascade decays also.

The resonance decays without transverse flow were investigated in
the paper~\cite{35}. We have calculated here the some resonance
decays spectra at SPS and RHIC energies. For example, spectra of
$\Lambda$ at decay $\Sigma^0 \to \Lambda\gamma$ is calculated
thus:
\begin{eqnarray}\label{eq.27}
&&\left(\frac{dN^{\Lambda}}{2 \pi
m_{\perp}dm_{\perp}}\right)_{y=0} =\\ \nonumber &=& \frac{V_{f}}{4
\pi^{3}}\frac{ m_{\Sigma} b g}{ p_1^{*} 4 \pi} \int\limits_0^1 dq
\frac{T_f^2}{m_{\perp}} e^{\frac{\mu_B - \mu_s}{T_f}}
f_{12}(p_{\perp},q,\bar v_{\perp}) q.
\end{eqnarray}
Here $b$ - branching, $g$ - stat.weight, $$ p_1^{*} =\frac{
\sqrt{[(m_{\Sigma}+ m_1)^2 - m_2^2][(m_{\Sigma} -m_1)^2 -
m_2^2]}}{2 m_{\Sigma}},$$ $m_1 = 1115$ MeV, $m_2 = 0$. We have
also:
\begin{equation}\label{eq.28}
f_{12} = 2 \int\limits_0^{z(p_{\perp})}dy
\frac{f_{11}(p_{\perp},y,q,\bar v_{\perp})
\cosh{(y)}}{\sqrt{\cosh^2{y} - (\frac{p_{\perp}}{m_{\perp}})^2}}.
\end{equation}
Here $z(p_{\perp}) =\ln(\frac{\sqrt{E_1^2 +p_{\perp}^2} +
p_1^*}{m_{\perp}})$, $E_1 = \frac{m_{\Sigma}^2 + m_1^2 - m_2^2}{2
m_{\Sigma}}$.

The value $f_{11}$ in formula (\ref{eq.28}) is:
\begin{eqnarray}\label{eq.29}
&&f_{11}(p_{\perp},y,q,\bar v_{\perp}) =
\int\limits_{x_1(p_{\perp},y)}^{x_2(p_{\perp},y)} dx x^2 e^{-x
\cosh{(y)} r(q,\bar v_{\perp})} \times \nonumber \\ &\times&
\frac{I_0(\sqrt{(x^2 - (\frac{m_{\Sigma}}{T_f})^2)(r^2(q,\bar
v_{\perp}) -1)}}{\sqrt{(x_2(p_{\perp},y)-x)(x-x_1(p_{\perp},y))}}.
\end{eqnarray}
Here~\mbox{$r(q,\bar v_{\perp})=1/\sqrt{1-(\frac{3}{2}q \bar
v_{\perp})^2}$}, \linebreak[4]\mbox{$x_{1,2}(p_{\perp},y) =
\frac{m_{\Sigma} (E_{1}m_{\perp}\cosh{(y)} \mp
p_{\perp}\sqrt{E_1^2 + p_{\perp}^2 -m_{\perp}^2
\cosh^2{(y)}}}{m_{\perp}^{2}\cosh^2{(y)} - p_{\perp}^2}$}.

At cascade decays, for example $\bar \Sigma^{0}\to \bar \Lambda
\gamma$, $\bar \lambda \to \bar p \pi$, we calculate at first the
spectra of $\bar \Lambda$, however no at angle of 90 degrees (i.e.
$y=0$), but for the whole y, and then this spectra we use for
calculation of spectra $\frac{dN^{\bar p}}{2\pi p_{\perp}
dp_{\perp}}$. In spectra $\frac{d\bar \Lambda} {2\pi
p_{\perp}dp_{\perp}}$ the value $\exp{\{-x\cosh{(y)}\; r(q,\bar
v_{\perp})\}}$ we replace now by the value $2K_{1}(x\; r(q,\bar
v_{\perp}))$. Then the spectra $\frac{dN^{\bar \Lambda}}
{m_{\perp}dm_{\perp}} \equiv f(m_{\perp})$ we substitute to
formula:
\begin{eqnarray}\label{eq.30}
&&\left(\frac{dN^{\bar p}}{2\pi p_{\perp}dp_{\perp}}\right)_{y=0}
= \frac{m_{\Lambda} b g_{\Lambda}}{4\pi p*}
\frac{2m_{\Lambda}}{\pi m_{\perp}^{\bar p}}\times \\\!\!\!
&\times& \!\!\!\!\!\!\int\limits_0^{z(p_{\perp})}\!\!\!
\frac{dy}{\sqrt{\cosh^2{(y_{\Lambda})-\frac{p_{\perp}^2}{m_{\perp}^2}}}}
\int\limits_{x_1(p_{\perp},y_{\Lambda})}^{x_2(p_{\perp},y_{\Lambda})}\!\!\!\!\!
dy_{1} \frac{y_{1}f(m_{\Lambda}\sqrt{y_1^2
-1})}{\sqrt{(x_2-y_1)(y_1-x_1)}}.\nonumber
\end{eqnarray}
We apply now to RHIC energy ($\sqrt{s}$=130 GeV).  At measurement
of antiprotons spectra in collaboration STAR were measured the
spectra of direct $\bar p$ and also spectra of $\bar p$ from weak
decays (with cascade). The our calculation shows that contribution
of weak decays is $\simeq 30\%$. We give here the meanings of
values, which were found for RHIC at $T_f=120$ MeV (by analogy
with SPS): $\mu_B^f$ =35.3 MeV, $\mu_s^f$=7.93 MeV, $n_f^{N-\bar
N}\simeq 1.062\times 10^{-3}m_{\pi}^3$, $s_B^f\simeq 0.073
m_{\pi}^3$, $s_{h+k}^f \simeq 1.28 m_{\pi}^3$.
  From conservation of
number net nucleons we have the volume $V_f$ at \mbox{$T_f = 120$
MeV} without accounting of weak decays: \linebreak[4]\mbox{$V_f =
\frac{N-\bar N}{n_f} \simeq 15250 m_{\pi}^{-3}$}. The entropy of
baryons in this volume is $S_B^f = V_{f} s_B^f \simeq 1114$. The
entropy of mesons is \mbox{$S_0 - S_B^f \simeq 5236$}. The mesons
entropy density is \linebreak[4]\mbox{$s_{h+k}^f \simeq 1.28
m_{\pi}^3$}. From here $V_{h+k}^f \simeq 4090 m_{\pi}^{-3}$.

\begin{figure} [ht]
\includegraphics*[scale=0.5]{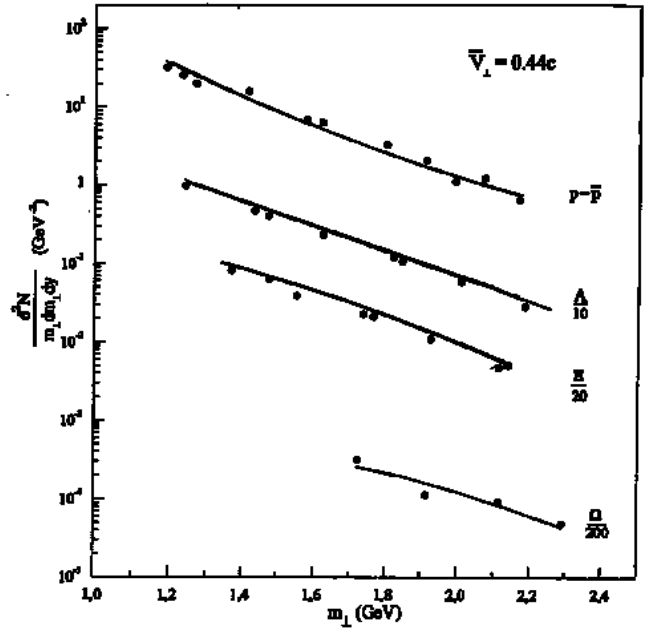}
\caption{The baryons spectra in Pb+Pb collisions at SPS energy in
central region of rapidity (quasiparticle model). The $\Lambda$ -
are  direct and with resonance decays:
$\Sigma^{1190}\to\Lambda\gamma$, $\Sigma^{1385}\to\Lambda\pi$,
$\Xi^{1315}\to\Lambda\pi$. The $\Xi^{1315}$ spectra - are  direct
+ $\Xi^{1530}\to\Xi\pi$ + $\Omega^{-}\to\Xi^{0}\pi^{-}$. The
spectra $p-\bar p$ and $\Omega_{1672}^{-}$ - are thermal direct.
The averaged transverse flow velocity $\bar v_{\perp} = 0.44c$.
 The experimental data are from~\cite{29,39}).
\label{Fig.1}}
\end{figure}

\setcounter{figure}{1}
\renewcommand{\thefigure}{\arabic{figure}a}

\begin{figure}
\includegraphics*[scale=0.5]{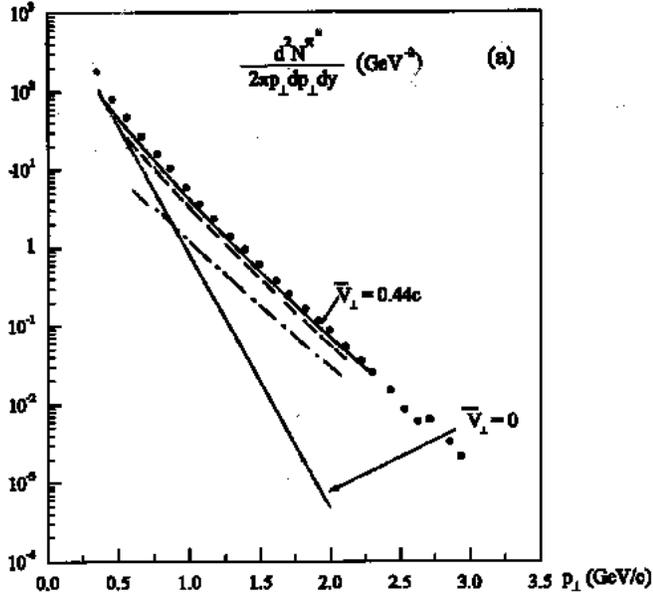}
\caption{The $\pi^{0}$
spectra (SPS, Pb+Pb) are direct + decays $\rho\to 2\pi$ +
$f_{2}\to 2\pi$ ($\bar v_{\perp} = 0.44c$). The dashed line - the
direct $\pi^{0}$, the dash-dotted line --- $\pi^{0}$ at decays. We
show also the summary spectra at $\bar v_{\perp} = 0$. The data
--- from~\cite{34}). \label{Fig.2a}}
\end{figure}

\setcounter{figure}{1}
\renewcommand{\thefigure}{\arabic{figure}b}

\begin{figure}
\includegraphics*[scale=0.5]{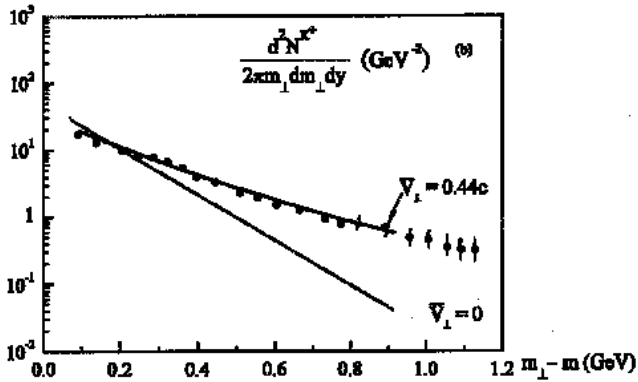}
\caption{The $k^{+}$ spectra (SPS,Pb+Pb) - direct + decay
$K_{890}^{*}\to k\pi$ ($\bar v_{\perp} = 0.44c$). We show also the
same spectra at $\bar v_{\perp} = 0$. The data ---
from~\cite{30}). \label{Fig.2b}}
\end{figure}

We have found the particles density at $T_f$:  $n_{\pi^{+}}^f
\simeq 0.067 m_{\pi}^3$, $n_{k^{+}}^f \simeq 0.0115 m_{\pi}^3$.
This gives the estimation of number of particles:
\linebreak[4]\mbox{$N_{\pi^{+}} \simeq 274$}, $N_{k^{+}} \simeq
47$. The value $V_f^{h+k}$ we use for calculation, for example, of
spectra $k^{-}$ and $\pi^{+}$ for energy RHIC. One can also to
show that relation of type (\ref{eq.24}) for thermal freeze-out
are fulfilled and for RHIC energy.

One can calculate the contribution of weak decays into net
nucleons density at $T_f = 120$ MeV (for RHIC)~\cite{33}:
$n_{w}\simeq 2.96\times 10^{-4} m_{\pi}^3$. With account of
nucleons from decays we find equivalent volume $V_f^{'}$:
\mbox{$n_{n-\bar N} V_f^{'}=16.3 + n_{w} V_f^{'}$}. From here
\linebreak[4]\mbox{$V_f^{'} \simeq 21150 m_{\pi}^{-3}$}.

The spectra of direct and of daughter $\bar p$ at weak decays we
calculate by formulas (\ref{eq.26})-(\ref{eq.30}) with factors
type $V_f^{'}\exp{\frac{-\mu_B^f}{T_f}}$, $V_f^{'}\exp{\frac
{-\mu_B^f +\mu_s^f}{T_f}} \cdots$. The contribution of weak decays
with cascade decays to $\bar p$ spectra gives $\simeq 30\%$.

\setcounter{figure}{2}
\renewcommand{\thefigure}{\arabic{figure}a}
\begin{figure}[ht]
\includegraphics*[scale=0.5]{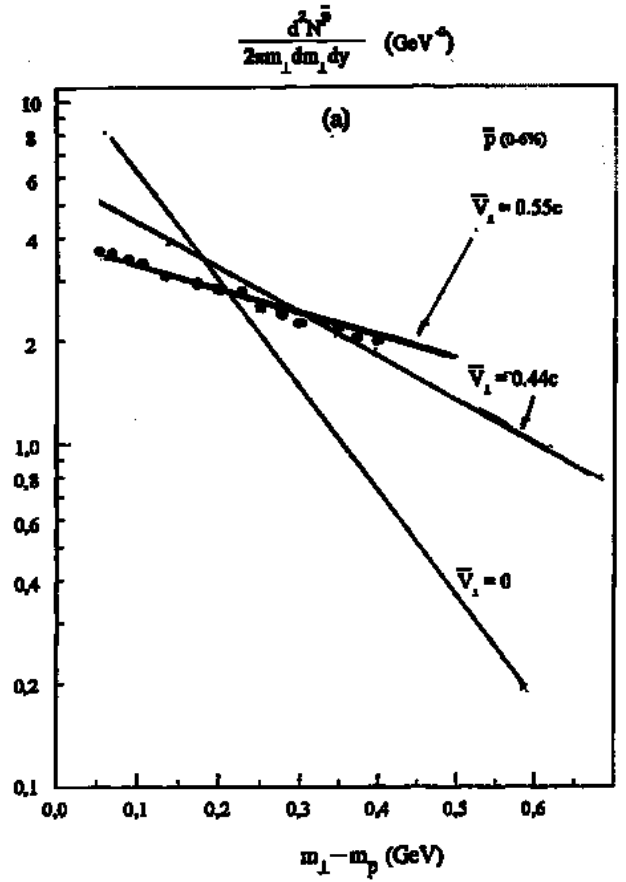}
\caption{The spectra $\bar p$ (centrality $0-6\%$) at RHIC energy
with accounting of weak decays($\bar v_{\perp} = 0.55c$). We show
also the same results with $\bar v_{\perp} = 0.44c$ and $\bar
v_{\perp} = 0$ (quasiparticle model). The data from~\cite{22}).
\label{Fig.3a}}
\end{figure}

\setcounter{figure}{2}
\renewcommand{\thefigure}{\arabic{figure}b}
\begin{figure}
\includegraphics*[scale=0.5]{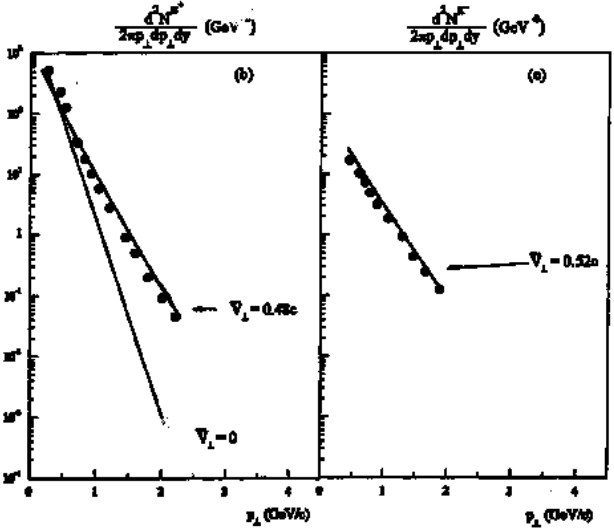}
\caption{The $\pi^{+}$ spectra (RHIC) - direct + decays $\rho\to
\pi\pi$ ($\bar v_{\perp} = 0.48c$). It is shown the same spectra
with $\bar v_{\perp} = 0$). \label{Fig.3b}}
\end{figure}
\setcounter{figure}{2}
\renewcommand{\thefigure}{\arabic{figure}c}
\begin{figure}
\caption{The $k^{-}$ spectra (RHIC) - direct + decays
$K_{892}^{*}\to k\pi$ for $\bar v_{\perp} = 0.52c$. The data ---
from~\cite{22,40}. \label{Fig.3c}}
\end{figure}

The theoretical $\bar p$ spectra  and experimental spectra for
central bin $0-6\%$ (STAR Coll.) at mid rapidity are shown in
Fig.~\ref{Fig.3a}. In  Fig.~\ref{Fig.3b}, Fig.~\ref{Fig.3c}
 we show also the direct
$\pi^{+}$ spectra and with decay $\rho \to \pi\pi$ and the $k^{-}$
spectra direct + $k_{892}^{*} \to k\pi$ and experimental data. In
Fig.~\ref{Fig.1} we show the baryons spectra: $p-\bar p$,
$\Lambda$, $\Xi$, and $\Omega$ for SPS (at $T_f$) in central
interval of rapidity. The $\pi^0$ and $k^{+}$ spectra are shown in
Fig.~\ref{Fig.2a}, Fig.~\ref{Fig.2b} together with data. One can
to see that spectra for RHIC have greater inverse slope than for
SPS. With increasing mass of particle the inverse slope (i.e. the
value $\bar v_{\perp}$) increases. Of course, one should note that
obtained spectra with sufficiently large value of $\bar v_{\perp}$
are correct only for not great values of $p_{\perp}$.

Thus, in the  present quasiparticle model with isentropic
evolution the normalization of baryons and mesons spectra do not
defined by free parameteres, but defined by conditions in initial
state ($T_0$, $V_0$), i.e. by quark-gluon plasma.

\section{The calculation by the use of  perturbative theory \label{sec5}}

We had investigated the physical characteristic of initial and
mixed phases and some baryons and mesons spectra for SPS and RHIC
in effective quasiparticle model with isentropic evolution. In
this section we shall consider at first the analogous problem of
nuclear collisions by the use of ordinary perturbative
decomposition of thermodynamic values in powers of running
coupling $\alpha_{s}(T)$ in the form~\cite{41}: $\alpha_{s}(T) =
6\pi/(11N_c-2N_f)\ln((2m_s+3T)/\lambda)$ up to order
$O(\alpha_{s})$ with QCD parameter $\lambda \sim 0.2$ GeV. Here
the value $Q^2$ in coupling $\alpha_{s}(Q^2)$ is considering as
the average of $Q^2$ in S-canal at temperature T. For example, the
entropy density for gluons in plasma is : \linebreak[4]\mbox{$s_g
= 32/45 \pi^2 T^3 (1 - \frac{15\alpha_{s}}{4\pi})$} and so on. We
calculate here also the values $T_0, V_0, S_0$.

The initial temperature $T_0$ can be find from formula analogous
(\ref{eq.23}), where we have now:
\begin{equation}\label{eq.31}
D_0(m_q, \mu_q, T_0, m_s, \lambda) = 1.
\end{equation}
Here the values $d_2(m_q, \mu_q, T_0, \lambda)$ and $n_{02}(m_q,
\mu_q, T_0, m_s, \lambda)$ are extended up to order $O(\alpha_s)$.
We find from here for \linebreak[4]\mbox{$\alpha = 1.1$}, $\mu_B =
247$ MeV, $m_q = 0$, $\lambda = 180$ MeV the such meanings(for
SPS): $T_0 = 178$ MeV, \mbox{$\alpha_{s}(T_0)=0.454$},
\mbox{$s_0(T_0)\simeq 26.4 m_{\pi}^3$}, \mbox{$s_c(T_c)\simeq
22.4$}, \mbox{$n_1(T_0)=0.475$}, \linebreak[4]\mbox{$n_1(T_C) =
0.426$} in units $m_{\pi}^3$,  $V_0 = N_0/n_1(T_0) \simeq 120
m_{\pi}^{-3}$, $S_0 = s_{0}V_0 = 3180$ (where $N_0\equiv (N - \bar
N)\simeq 57$).

In this model the entropy of nucleons and strange baryons is the
same as in quasiparticle model: $S_B^f = 992$, the entropy of
mesons \mbox{$S_{h+k}^f = S_0 - S_B^f \simeq 2200$} and
\linebreak[4]\mbox{$V_{h+k}^f \simeq 1700 m_{\pi}^{-3}$}. We have
here $N_{\pi^0}^f \simeq$ 136, $N_{k^{+}}\simeq$ 25.

For RHIC energy we find for $\alpha = 1.11$, $\mu_B = 50$ MeV,
$m_q = 0$, $\lambda = 180$ MeV the meanings: $T_0 \simeq 219$ MeV,
$\alpha_s(T_0) = 0.418$, $s_{0}(T_0) = 50.11 m_{\pi}^3$,
$n_1(T_0)=0.145$, \linebreak[4] \mbox{$n_1(T_c)= 0.084
m_{\pi}^3$}, $V_0 = N_0/n_{1}\simeq 111.7, V_c \simeq 193
m_{\pi}^{-3}$, $S_0 = s_0 V_0 \simeq 5600$. We have here
\mbox{$\tau_0 \simeq 2.16$ fm}, \linebreak[4]\mbox{$\tau_c \simeq
3.73$ fm}.
 One can  show, that in constituent quarks phase the relations~(\ref{eq.24})
here also are fulfilled only with decrease of number of degrees of
freedom in the presence of pseudogoldstone states.

We have here by analogy with quasiparticle model $V_f \simeq 15250
m_{\pi}^3$, the entropy of baryons $S_B^f \simeq 1114$, the
entropy of mesons is ~\mbox{$S_{h+k}^f = S_0 - S_B^f = 4486$} and
\linebreak[4]\mbox{$V_{h+k}^f = \frac{S_{h+k}^f}{s_{h+k}^f} =3500
m_{\pi}^{-3}$ ($s_{h+k}^f \simeq 1.28 m_{\pi}^3$)}. That gives the
estimation: $N_{\pi^{+}}\simeq 243$, $N_{k^{+}}\simeq 40$.

In  this simple model we have not great decrease of normalization
of mesons spectra at SPS and RHIC \linebreak[4] (by 12-13 \%) in
comparison with quasiparticle model, i.e. apparently it takes
place within the limits of precision of measurement. Thus we do
not have here noticeable difference for spectra of particles in
comparison with quasiparticle model.
However this model disagrees
with $SU(3)$ lattice data in region of phase transition. The
values $s/T^3$, $\epsilon/T^4$, $p/T^4$ decrease very weakly in
this case (For example, at $3T_c \to T_c$ the value $s/T^3$
decrease only by value $\simeq 0.78$) The coupling strength
$\alpha_{s}(T)$ increase(although not great) at $T\to T_c$ from
above (unlike effective quasiparticle model). Thus we see, that
spectra of particles weakly depend on character of phase
transition $QGP \to$ hadrons.

But one can show~\cite{12}, that in this model we have too great
energy loss $\Delta E_g$ of gluon jet at RHIC energy: $\sim
80-90\%$. That is caused in the main by  value of effective
coupling strength.

Let us consider now the quasiparticle model with phenomenological
parametrization of running coupling $G(T)$ (\ref{eq.8}) We use
here the formulas of type (\ref{eq.10})-(\ref{eq.13}) and of type
(\ref{eq.14})-(\ref{eq.18}) for calculation of thermodynamic
values $m_q$, $s_0$ and $n_0$, but with coupling $G(T)$
(\ref{eq.8}) and with $C(T,T_c)$=1. With help of corresponding
parameters $T_s$, $\lambda$ it is possible well to fit the new
lattice data even near of $T_c$. For calculation of initial
conditions we use the equation of type (\ref{eq.23}). For example,
for possible parametrization ~\cite {6}: $\lambda$=5.3,
$T_s/T_c$=0.73 (at $N_f$=3) we find for RHIC energy
($\sqrt{s}$=130 GeV) the following meanings: $T_0 \simeq$ 250
MeV,$G(T_0) \simeq$ 2.533 (i.e. $\alpha_{s}(T_0)\simeq$ 0.509) and
also $\alpha_{s}(T_c) \simeq $ 1.948. We find corresponding values
for entropy and net nucleons density: $s_0(T_0)\simeq 84.36
m_{\pi}^3$, $s_c(T_c)\simeq 12.96 m_{\pi}^3$, $n_1(T_0)\simeq
0.210 m_{\pi}^3$, $n_1(T_c)=0.0442 m_{\pi}^3$. From here we have
the estimation for initial state: $\tau_0 \simeq(N-\bar
N)/(n_1(T_0)\pi R_{Au}^2) \simeq 1.5$ fm., and by analogy $\tau_c
\simeq 7.14$ fm. We have for complete entropy: $S_0 V_0 =6548$,
i.e. only on $\sim 3\%$ the difference from its value in effective
quasiparticle model. Thus we have practically identical baryons
and mesons spectra  in these two models.

However we have calculated the jet quenching in plasma in this
model by analogy with effective quasiparticle model~\cite{12}. The
result is dramatic: the energy loss $\Delta E_g$ of gluon jet (for
two parametrization of $\lambda$ and $T_s/T_c$) four - five times
as much than energy $E_g$ of itself jet !. This contradiction is
caused by too great value of running coupling $G(T)$(\ref{eq.8})
in this model, which besides increase at $T \to T_c$ from above.

\section{Conclusion \label{sec7}}

In this paper we investigate the isentropic evolution of expanding
quark-gluon plasma  with  phase transition at first into plasma
part and then hadron part of mixed phase at temperature $T_c$, and
further isentropic expansion up thermal freeze-out at temperature
$T_f$. For investigation of initial plasma phase we use the
quasiparticle model, which allow apparently to sum up partially
the perturbative expansion of thermodynamic function of
interacting plasma in powers of coupling constant. In the
papers~\cite{4,6} was used the phenomenological parameterization
of coupling constant in quasiparticles model in accordance with
lattice data. However close to $T_c$ the coupling constant is
sufficiently great, therefore in the vicinity of the phase
transition the perturbative methods are not expected to be
reliability .

In the paper~\cite{10} was shown, that at such parameterization
the thermal mass of gluon grows in the vicinity of the phase
transition when $T \to T_c$ from above. This do not agrees with
SU(3) lattice data for Debye mass. In the paper~\cite{10} was
considered the phenomenological model of confinement, where the
decrease of thermodynamic values at $T\to T_c$ from above is
caused no increase of mass, but by decrease of the number of
active degrees of freedom (i.e. by modification $g_g \to C(T)
g_g$). It is supposed, that behaviour of thermal gluon mass
$m_g(T)$ is similar with behaviour of Debye mass, i.e. it
decreases with $T \to T_c$ from above. This gives the good
description of SU(3) lattice data. The similar modification is
used also for quasiparticle model with quarks.

It is assumed that Debye gluon mass have form
\linebreak[4]\mbox{$m_{D}(T) = \tilde G(T)T$} and it has a small
gap at $T = T_C$ in accordance with SU(3) lattice data. It is
assumed also that the proportionality of type (\ref{eq.5}) between
$m_{D}$ and thermal gluon mass $m_g$ remains true in the vicinity
of phase transition.

We use such model for description of isentropic evolution in
common case of quasiparticles with $\mu_{B}\ne 0$ for calculation
of  physical characteristic of initial plasma phase: $T_0$  and
$V_0$. In result we have obtained the formula (\ref{eq.23}) for
definition of these values, where we use the approximation
(\ref{eq.20}) for $\left(\frac{dN}{dy}\right)_{A+A}$ and
experimental meaning of average number of net nucleons $(N -\bar
N)$ in central region of rapidity at SPS and RHIC energies. From
conservation of entropy and of number of net nucleons follow also
the relations (\ref{eq.24}), which must be fulfilled
simultaneously with (\ref{eq.23}). The relations (\ref{eq.23}),
(\ref{eq.24}) depend on effective coupling constant $G_0$, which
is determined by asymptotic value of thermal gluon mass at $T =
3T_c$~\cite{13}. For that we use the asymptotic value of running
coupling $\alpha_{s}(T)$~\cite{6}. In result were obtained the
initial meaning $T_0$, $V_0$ at SPS and RHIC energy: for SPS we
have $T_0 \simeq 175$ MeV, $V_0 \simeq 518$ fm$^3$, (i.e. $\tau_0
\simeq 3.28$ fm) and $V_c \simeq 650$ fm$^3$ (i.e.$ \tau_c \simeq
4.1$ fm), and for RHIC $T_0 \simeq 219.6$ MeV, $V_0 \simeq 322$
fm$^3$ and $V_c \simeq 965$ fm$^3$, i.e. $\tau_0 \simeq 2.2$ fm,
$\tau_c \simeq 6.5$ fm. We have also $S_0 \simeq 3508$ and $\simeq
6350$ for SPS and RHIC correspondingly.

When the temperature is lowered ($T\to T_c$), more and more gluons
form heavy glueballs, and the weaker becomes the interaction. This
state can be interpreted as superposition of nonperturbative
multigluon and weak glueball exchange~\cite{10}. The interaction
between the quarks also becomes weaker. It is possible to
interpret, that there is appearance of massive constituent quarks,
while the multigluon state is absorbed into constituent quarks
\linebreak[4] masses.

Further we show, that massive constituent quarks appears with
decrease of number of degrees of freedom in the presence of octet
of pseudogoldstone states. That follow from conservation of
entropy and of number of net nucleons.  The hadron part of mixed
phase appears also with decrease of number of degrees of freedom
--- the same as for constituent quarks. However at thermal
freeze-out (at $T=T_f$) hadrons appears already with normal number
of degrees of freedom.

Thus we have at SPS the short plasma phase. With above-mentioned
parameters we have calculated in effective quasiparticle model the
some typical baryons and mesons spectra with account of resonance
decays. We use for heavy nuclei the temperature of thermal
freeze-out $T_f = 120$ MeV (as in  "standard" model~\cite{31,32}
The normalization of baryons and mesons spectra in the present
model is no free parameter, but defined by initial condition in
quark-gluon plasma.

One should note that there are the models for description of
particle production with simultaneous chemical and thermal
freeze-out~\cite{36}, where 2-3 free parameters describe well the
spectra at RHIC energy. In the present work we use "standard"
scenario with two freeze-out - chemical and thermal.

The main object of present work is no study the whole spectra at
SPS and RHIC, but the study of dependence of spectra from initial
state and from phase transition stage. For this aim we consider in
Sec.\ref{sec5} the analogous isentropic problem by the use of
ordinary perturbative theory up to order $O(\alpha_{s}(T))$, where
$\alpha_{s}(T)$ is running coupling. We do not find in this
approximation of noticeable difference for particles spectra in
comparison with effective quasiparticle model, although  this
model disagrees with SU(3) lattice data in region of phase
transition (unlike quasiparticle model). However it can be
shown~\cite{12} that in this model we have the too great  energy
loss of gluon and quark jets in plasma ($\sim 80-90\%$), that
contradict to suppression of $\pi^0$-spectra, reported by PHENIX
~\cite{40}.

In this section we consider also the quasiparticle model with
phenomenological  parametrization of coupling $G(T)$, which gives
good fit of new lattice data. We investigate here also the initial
condition in plasma at RHIC energy (on the basic of equation of
type(\ref{eq.23}). We find that baryons and mesons spectra in this
model practically do not differ from spectra in effective
quasiparticle model.

However the situation with energy loss in plasma here yet more
dramatic - the energy loss of gluon jet exceed the energy of
itself jet. Thus apparently we have correct quantitative
description of energy loss of gluon and quark jets in plasma only
in considering  here effective quasiparticle model.

But of course the question about nonperturbative solution for
thermodynamic values in quark-gluon plasma remains open (for
example, for thermal mass of quasiparticles and running coupling
and so on). Unfortunately in the vicinity of the phase transition
is used the phenomenological parameterization for these values.

In this model we do not consider also  the intrinsic volume of
particles. We had investigated in the main the ratios of
thermodynamic values. We use the models~\cite{37,38}, where
effects due to excluded volume correction cancel out in ratios.

At calculation in central region of rapidity we do not take into
account the difference of number protons and neutrons.

It should be noted also the decrease of protons suppression for
moderate $p_{\perp}$ at RHIC energy~\cite{40} It is possible to
use for this problem the hydrodynamic spectra in quasiparticle
model for $p$ and $\bar p$ with accounting of resonance decays.

The author thank S.T.Beljaev, V.I.Manko and L.M.Satarov for
fruitful discussion. The work was supported by grant
NS-1885.2003.2.

\end{document}